\documentclass{jpsj3}
\usepackage{txfonts}

\newcommand {\rPHE}{\rho_{yx}^{\rm{PHE}}}
\newcommand {\Hsko}{H_{\rm{sk1}}}
\newcommand {\Hskt}{H_{\rm{sk2}}}
\newcommand {\Hc}{H_\mathrm{c}}
\newcommand {\Tco}{T_{\mathrm{c}1}}
\newcommand {\Tct}{T_{\mathrm{c}2}}

\newcommand {\Tc}{T_{\mathrm{c}}}

\newcommand {\rphe}{\rho_{yx}^{\rm{PHE}}}
\newcommand {\rpara}{\rho^{\parallel}}
\newcommand {\rperp}{\rho^{\perp}}
\newcommand {\rparam}{\rho^{\parallel {\bm M}}}
\newcommand {\rperpm}{\rho^{\perp {\bm M}}}
\newcommand {\rparaq}{\rho^{\parallel {\bm Q}}}
\newcommand {\rperpq}{\rho^{\perp {\bm Q}}}
\newcommand {\rPHEM}{\rho_{yx}^{{\rm PHE},\, \bm{M}}}
\newcommand {\rPHEQ}{\rho_{yx}^{{\rm PHE},\, \bm{Q}}}
\usepackage{graphicx}
\usepackage{bm}
\usepackage{pifont}
\usepackage{amsmath}

\title{Formation of In-plane Skyrmions in Epitaxial MnSi Thin Films as Revealed by Planar Hall Effect}

\author{T. Yokouchi$^1$\thanks{yokouchi@cmr.t.u-tokyo.ac.jp}, N. Kanazawa$^1$, A. Tsukazaki$^2$, Y. Kozuka$^1$, A. Kikkawa$^3$,\\Y. Taguchi$^{3}$, M. Kawasaki$^{1,3}$, M. Ichikawa$^1$, F. Kagawa$^{3}$, and Y. Tokura$^{1,3}$}
\inst{$^1$Department of Applied Physics and Quantum Phase Electronics Center (QPEC), University of Tokyo, Tokyo 113-8656, Japan \\
$^2$Institute for Materials Research, Tohoku University, Sendai 980-8577, Japan \\
$^3$RIKEN Center for Emergent Matter Science (CEMS), Wako 351-0198, Japan} 

\abst{We investigate skyrmion formation in both a single crystalline bulk and epitaxial thin films of MnSi by measurements of planar Hall effect. A prominent stepwise field profile of planar Hall effect is observed in the well-established skyrmion phase region in the bulk sample, which is assigned to anisotropic magnetoresistance effect with respect to the magnetic modulation direction. We also detect the characteristic planar Hall anomalies in the thin films under the in-plane magnetic field at low temperatures, which indicates the formation of skyrmion strings lying in the film plane. Uniaxial magnetic anisotropy plays an important role in stabilizing the in-plane skyrmions in the MnSi thin film.}

\begin{document}
\maketitle

\section{Introduction}

 Magnetic skyrmions in chiral magnets are spin-swirling vortex-like matters as topologically defined by an integer winding number, and hence show versatile emergent electromagnetic responses \cite{review}. Because of possible electrical controls of skyrmions, such as ultralow current-density drive \cite{Jonietz,Yu3}, electric-field induced motion \cite{White}, and read/write operations by spin-polarized currents \cite{Romming}, skyrmions are considered as a promising candidate of the information carrier in emerging spintronics \cite{Sampaio,Koshibae}. Forms of skyrmion aggregate or skyrmion crystal in confined geometries of chiral magnets, including thin films \cite{Karhu,Huang,Yokouchi}, nanowires \cite{Kanazawa} and nano areas \cite{Park},  are of particular interest in the light of enhanced stability of skyrmions with higher density. 

Magnetic phase diagrams for the skyrmion-hosting bulk materials with the common space group $P2_1 3$ share a universal profile \cite{Bauer} as shown in Fig.~1(a). While the skyrmion state in bulk is  stabilized only near the transition temperature ($\Tc$) by thermal fluctuation \cite{Muhlbauer}, the skyrmion phase extends over a wide temperature ($T$)-magnetic field ($H$) region in case of thin films with a magnetic field perpendicular to the film plane \cite{LTEMFeCoSi}. The geometrical effect suppresses the formation of conical structure [Fig.~1(d)] modulating along the magnetic field direction (the out-of-plane direction); consequently, periodically-arranged skyrmions in the plane normal to $H$ [Fig.~1(e)] become a globally stable state in the thin films. A systematic real-space observation on freestanding MnSi thin plates with thickness gradients and different crystalline orientations has revealed that the relative skyrmion's stability against the film thickness in the respective crystallographic planes of the films is determined by competition among Dzyaloshinskii-Moriya interaction, dipole-dipole interaction, and uniaxial magnetic anisotropy  \cite{Yuetal}.

However, until now there have been few studies on an effect of uniaxial magnetic anisotropy on stability of skyrmion, which effect was first proposed for explaining the enhanced skyrmion phase in the thin films \cite{Butenko}. Here, to gain insight into this effect, we focus on one other type of skyrmion aggregate, that is, an array of skyrmion rows stretching in the plane of strained MnSi thin films with an in-plane $H$ \cite{Wilson}. Epitaxial MnSi(111) thin films on Si substrates receive a tensile strain due to the lattice mismatch, which increases the hard-axis uniaxial anisotropy along the direction normal to the film plane \cite{Wilson2}. Detection of such an in-plane skyrmion formation, however, is challenging experimentally because the established detection methods, such as Lorentz transmission microscopy (TEM) and topological Hall effect \cite{Lee,Neubauer}, are difficult to be applied, in principle, for an in-plane $H$ configuration. Thus far, the formation of the in-plane skyrmion [Fig.~3(a)] in the thin films has been proposed only from magnetization measurements \cite{Wilson}.

In this paper, we demonstrate a new detection method for the formation of the in-plane skyrmion strings appearing in a thin film. By measurements of planar Hall effect (PHE), which sensitively extracts an anisotropic component of electrical conductance, we identify the emergence of skyrmions as a prominent stepwise field profile in the PHE signal for both a single-crystalline bulk and epitaxial thin films of MnSi. A $T$-$H$ phase for the in-plane skyrmions appears at low temperatures, which is distinct from the hitherto known skyrmion phase stretching from $\Tc$. The in-plane skyrmion strings are stabilized by the magnetic anisotropy, which is enhanced at low temperatures.

\begin{figure}
\begin{center}
\includegraphics*[width=8.5cm]{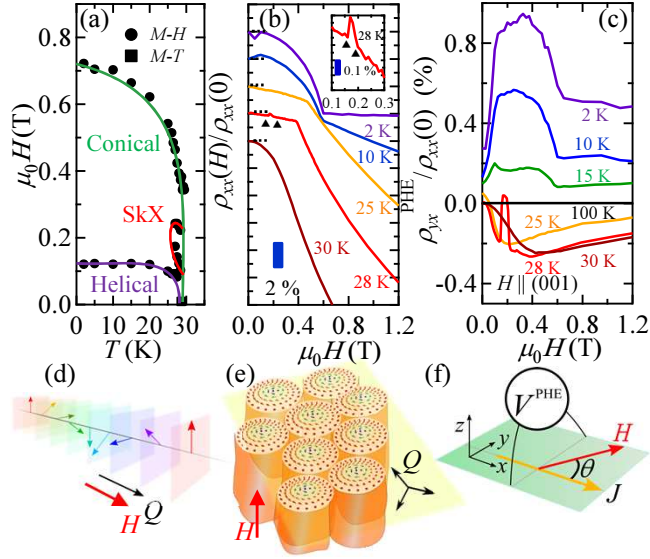}
\caption{(color online). (a) Magnetic phase diagram for the bulk crystal of MnSi determined by magnetization measurements. Magnetic-field dependence of (b) magnetoresistivity and (c) planar Hall resistivity in the bulk sample. The inset of panel (b) is a magnified image of magnetoresistivity at the skyrmion phase. Schematic illustrations of (d) conical structure and (e) skyrmion crystal. (f) Experimental setup for the measurement of PHE.}
\end{center}
\end{figure}

\section{Experiments}
The MnSi single crystal was grown by the Czochralski method and cut into rectangular shape with a typical size of $2\times1\times0.3$  $\rm{mm}^{3}$. The MnSi epitaxial films were grown on a Si (111) substrate by solid phase epitaxy as detailed in Ref. 10. Planar Hall effect is measured with a setup shown in Fig.~1(f). Magnetic field is applied in the $x$ (current direction)-$y$ (voltage direction) plane. Measured planar Hall resistivity $\rphe$ reads 
\begin{equation}
\rphe=\frac{1}{2}(\rpara-\rperp)\sin2\theta,
\end{equation}
where $\theta$ is angle between the current ($J$) and the magnetic field, $\rpara$ and $\rperp$ are resistivities with the current parallel and perpendicular to the magnetic field, respectively. 
Note that PHE originates from the anisotropic magnetoresistivity, not the conventional Hall effect. Because a dominant contribution to anisotropic magnetoresistance in ferromagnetic $3d$-transition-metal alloys \cite{AMR} is usually related to the magnetization ($\bm{M}$) direction, we use the magnetization vector as a reference direction for PHE measurements. Partly because the magnetization direction is parallel with the magnetic field in MnSi as well as in most ferromagnets, we generally equate Eq. (1) with the following relation: $\rPHEM=\frac{1}{2}(\rparam-\rperpm)\sin2\theta_{\bm M}$, where $\rparam$, $\rperpm$, and $\theta_{\bm M}$ are corresponding parameters measured with reference to ${\bm M}$. To remove voltages from Hall effect and longitudinal resistivity due to misalignments of the sample mounting and the electrodes, we measured the transversal voltage for $\pm H$ and $\pm\theta$ and then symmetrized it against $H$ and antisymmetrized it against $\theta$. Hereafter we define $\rPHE$ as its signal at $\theta = 45^{\circ}$ unless otherwise noted.

\section{Results and Discussions}
We first demonstrate that the PHE is a sensitive probe for identifying skyrmion formation through measurements on a well-studied skyrmion material, the MnSi single-crystalline bulk sample.
Figure~1(b) shows the $H$-dependence of magnetoresistivity $\rho_{xx}(H)/\rho_{xx}(0)$ at various temperatures for a setup of  $H\parallel J \parallel [110]$. The magnetoresistivity (MR) shows an inflection at the critical field $\Hc$, where the transition occurs between conical and ferromagnetic structures. In the magnetic field scan crossing the skyrmion phase, a small kink (0.1 \% change) in MR is also observed [inset of Fig.~1(b)], which is consistent with previous reports \cite{MR_Date,Demishev}. 
We compare planar Hall signals at the corresponding temperatures measured with $J \parallel [110]$ and $H$ lying in $(001)$ plane in Fig.~1(c). $\rPHE$ exhibits clear changes at the magnetic phase boundaries [see also Fig.~1(a)]. In particular, $\rPHE$ displays a distinctive stepwise anomaly at the skyrmion phase, which enables us to use $\rPHE$ as a sensitive probe for the skyrmion phase. Here we again note that the step-like behavior of PHE in SkX is not a contribution from THE because the symmetrization against $H$ removes Hall contribution as mentioned above; in fact the magnitude is approximately ten times larger than THE in MnSi \cite{Neubauer} [see also Fig.~2(c)]. To build further assurance about the correspondence between the skyrmion phase boundaries and $\rPHE$ anomalies, we present development of $\rPHE$ in the $T$-$H$ region around the skyrmion phase in Fig.~2(a). Sharp stepwise structures are confirmed between 27.0--28.5 K. In Fig.~2(b), we map the $H$-derivative of PHE [inset of Fig.~2(b)], which emphasizes the abrupt change in PHE, for comparison with the established phase diagram. The abrupt rises and falls of $\rPHE$ coincide with the phase boundaries determined by the magnetization measurements in the $T$-$H$ plane, from which we confirmly assign the PHE anomaly to the skyrmion formation.

The PHE anomaly at the skyrmion phase can be accounted for with a following phenomenological model. Provided that resistivity in a periodically modulated magnetic texture also depends on the orientation of the modulation vector ($\bm{Q}$), an additional contribution will appear obeying the following relation in a similar way to the conventional PHE with reference to the magnetization:  $\rPHEQ=\frac{1}{2}(\rparaq-\rperpq)\sin2\theta_{{\bm Q}}$, where $\rparaq$, $\rperpq$, and $\theta_{\bm Q}$ are corresponding parameters measured with reference to ${\bm Q}$.
Indeed, a recent study on anisotropic magnetoresistance (AMR) associated with the helical structure in $B$20-type (Fe, Co)Si has revealed that the magnetic modulation itself affects the electrical conductance, resulting in the difference between $\rparaq$ and $\rperpq$ \cite{Huang2}. Upon the transformation to the skyrmion state, $\rPHEQ$ changes its sign due to the sign inversion of $\sin2\theta_{\bm Q}$ accompanied by the 90$^{\circ}$-flop of ${\bm Q}$, which causes the distinctive anomaly. We note that the magnetic-field dependence of $\rPHE$ with passing through other magnetic phases [Fig.~1(c)] can be also explained on the basis of this phenomenological model: While the formation of a multidomain state of the single-${\bm Q}$ helical structure nearly cancels out $\rPHEQ$, the AMR feature is restored by $H$-alignment of the domains of the helical (conical) structure, as the enhanced absolute value of $\rPHEQ$ in the conical phase. When the ferromagnetic state is induced above $\Hc$, the contribution from $\rPHEQ$ disappears, leading to the reduction of $\rPHE$ magnitude.

The phenomenological expression is further verified by the angular dependence of PHE. 
Figure~3(a) shows PHE signals normalized by $\sin2\theta$ at various $\theta$ measured with the same setting for Figs.~1(c) and 2(a), i.e., $J\parallel [110]$ and $H\parallel (001)$. Since the spin $\bm{Q}$ vectors of the conical and skyrmion structures are parallel and perpendicular to $H$, respectively, each $\rPHEQ$ as well as $\rPHEM$ obeys the $\sin2\theta$ dependence. The angles between the electric current and magnetic modulation direction ($\theta_{\bm Q}$) become $\theta$ and $\theta +90^{\circ}$ in the conical and skyrmion phases, respectively. Consequently, the angle dependencies of PHE remain $\sin 2\theta$ in the both phases: $\sin2\theta_{\bm Q}=\sin2\theta$ and $\sin2\theta_{\bm Q}=\sin2(\theta+90^{\circ})=-\sin2\theta$. In fact, the signals of PHE normalized by $\sin2\theta$ trace the identical curve [Fig.~3(a)]. This is further confirmed by $\theta$ dependence of $\rPHE$ [Figs.~3 (d)-(e)]; planar Hall signals at each magnetic phase clearly follow $\sin2\theta$ curves. The same angular dependence of $\rPHE$ is confirmed also in different settings of magnetic field and crystallographic orientation [Figs.~3(b) and 3(c)], although they show much different $H$-profiles. The AMR ratio, i.e., the difference between $\rparaq$ and $\rperpq$, in the $B$20 compounds largely depends on the complex combination of anisotropic nature of scattering processes and band structure \cite{Helix_kang}, which probably causes the significant difference in both the magnitude and sign of $\rPHE$ as observed.

\begin{figure}
\begin{center}
\includegraphics*[width=8.5cm]{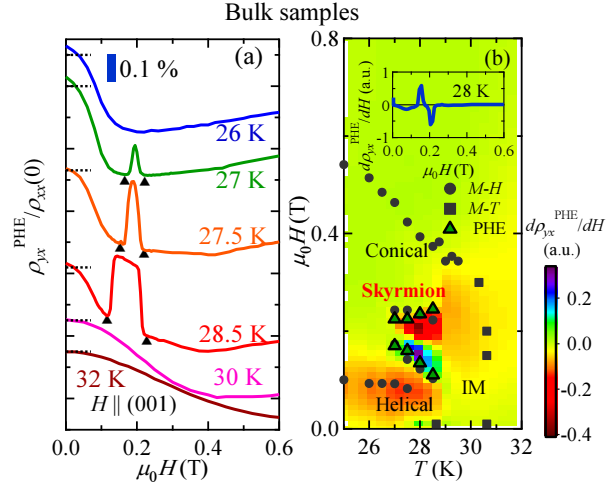}
\caption{(color online). (a) Magnetic-field dependence of planar Hall resistivity normalized by longitudinal resistivity at zero field around the skyrmion phase in the bulk sample. (b) A contour map of $H$-derivative of planar Hall resistivity. The solid circles and squares represent phase boundaries determined by magnetization measurements and the open triangles represent the points where the kinks of planar Hall resistivity are observed, corresponding to solid triangles in panel (a). The inset of panel (b) shows the $H$-derivative of planar Hall resistivity.}
\end{center}
\end{figure}

\begin{figure}
\begin{center}
\includegraphics[width=8.5cm]{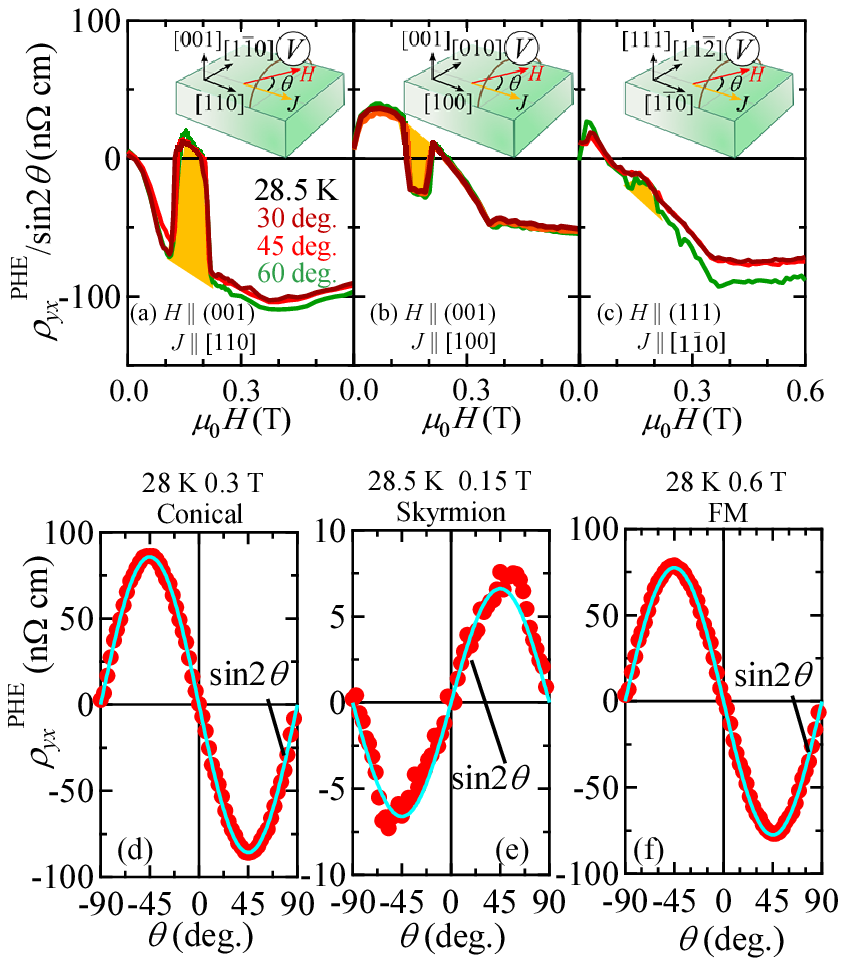}
\caption{(color online). Magnetic-field dependence of planar Hall resistivity normalized by $\sin2\theta$ at various $\theta$ with (a) $H $ lying in (001) plane and $J\parallel[110]$, (b) $H $ lying in (001) plane and $J\parallel[100]$, and (c) $H $ lying in (111) plane and $J\parallel[1\bar{1}0]$, respectively. The insets of panels (a)-(c) are experimental setups for the measurement of PHE. Angular ($\theta$) dependence of PHE in (d) conical phase, (e) skyrmion phase, and (f) ferromagetic (FM) phase with $H $ lying in (001) plane and $J\parallel[110]$. Here, $\theta$ is angle between the current and the magnetic field. The light blue lines are fits to $\sin2\theta$.}
\end{center}
\end{figure}

We apply the PHE measurement to detection of the in-plane skyrmion strings forming in the epitaxial MnSi thin film.  In Fig.~4 are presented the magnetic field dependencies of magnetization $M$, magnetoresistivity normalized by its value at zero field $\rho_{xx}(H)/\rho_{xx}(0)$, and PHE signal normalized by the longitudinal resistivity at zero field $\rPHE/\rho_{xx}(0)$, at three temperatures (2, 10, 30 K). Magnetoresistivity and PHE are measured with electric current $J\parallel [1\bar{1}0]$ and with magnetic field $H\parallel J$ and $H\parallel (111)$ surface, respectively.
It is obvious that PHE signal shows a distinctive anomaly characteristic of the skyrmion formation at low temperatures below 20 K [Figs.~4(i) and 4(j)]. Given the theoretical prediction \cite{Wilson}, the skyrmion strings stretching along the in-plane $H$ in the thin film are likely responsible for the PHE anomalies, as schematically shown in Fig.~4(a). Between 20 K and 40 K($\approx \Tc$), all the three quantities [$M$, $\rho_{xx}(H)/\rho_{xx}(0)$, and $\rPHE/\rho_{xx}(0)$] indicate only one distinct magnetic transition at $H_{\rm{c}}$ as exemplified in Figs.~4(e), 4(h), and 4(k). Above $T_{\rm{c}}$, no significant signals are observed (not shown).  We note that there are observed tiny anomalies in $M$ and $\rPHE/\rho_{xx}(0)$ at intermediate fields between the zero field and the critical field $\Hc$ at $T=$ 25--35 K [see also Figs.~4(e) and 4(k)]. These may indicate sparse formation of skyrmion strings.

The magnetic field range of the PHE anomaly ($H_{\mathrm{sk1}} < H < H_{\mathrm{sk2}}$) extends well above $H_{\rm{c}}$ [Figs.~4(i) and 4(j)] and even reaches zero field in the decreasing field process at 10 K [Fig.~4(j)]. Once skyrmions are created, they coexist with other magnetic phase persisting beyond their thermodynamical-stability $H$-region. This originates from the first-order phase transition nature associated with topological change in the magnetic texture, i.e., unwinding the skyrmions costs a considerable barrier energy. Because of the topologically-stable nature of skyrmions, the hysteretic skyrmion formation with respect to magnetic field change also shows up as the hysteresis in the PHE signal [Figs.~4(i) and 4(j)]. The PHE anomaly is more prominent in the course of increasing field than decreasing field at 2 K [Fig.~4(i)]. Since the magnitude of the PHE anomaly should be  associated with the skyrmion density, the large hysteresis in PHE indicates that the density of packed skyrmion strings depends on the precedented magnetic structure determined by the magnetic field history; the helical structure is more prone to the development of the skyrmions than the ferromagnetic state. With a slight elevation of temperature from 2 K, for example at 10 K, skyrmion formation occurs in different $H$-ranges between the increasing and decreasing field processes [Fig.~4(j)]. With increasing field, the transformation of the in-plane skyrmion strings from the helical structure occurs at $H_{\mathrm{sk1}}$, followed by the continued existence of skyrmions well above $\Hc$; with decreasing field, skyrmions appear at $\Hc$, remaining even near zero field. Here we note that while there are also discerned kinks and/or hysteretic behaviors corresponding to the skyrmion phase in the magnetization and magnetoresistivity curves, the planar Hall signal shows much better sensitivity for the skyrmion formation.   

\begin{figure}
\begin{center}
\includegraphics*[width=8.5cm]{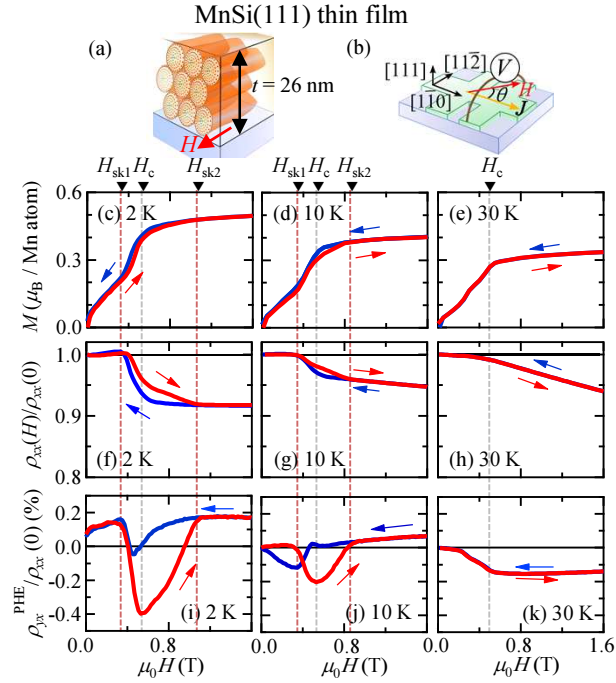}
\caption{(color online). (a) A schematic illustration of the skyrmion formation in the presence of in-plane magnetic field. (b) An experimental setup for the measurement of PHE.  Magnetic-field dependence of (c)-(e) magnetization, (f)-(h) magnetoresistivity, and (i)-(k) planar Hall resistivity of 26-nm MnSi thin film at 2 K, 10 K, and 30 K. Red lines indicate the data taken with increasing field and blue lines the data with decreasing field. The vertical dashed lines represent $\Hsko$, $\Hskt$, and $\Hc$; $\Hsko$ and $\Hskt$ correspond to the lower and upper critical fields of the $\rPHE$-hysteretic regime, where the $\rPHE$ originating from the in-plane skyrmions appears, and $\Hc$ stands for the critical field above which the spin collinear ferromagnetic state shows up.}
\end{center}
\end{figure}

We show contour mapping of $\rPHE/\rho_{xx}(0)$ for the increasing field process in Fig.~5(a), along with phase boundaries determined by measurements of $M$ and PHE. 
In contrast to the skyrmion phase in the bulk MnSi as stabilized by the large thermal fluctuations near $\Tc$, the in-plane skyrmion phase for the thin film appears at low temperatures; this indicates a different driving force is involved in the formation of the in-plane skyrmions. The uniaxial magnetic anisotropy enhanced at low temperatures is perhaps the major contribution as theoretically suggested \cite{Wilson}.
To highlight the hysteretic formation of the in-plane skyrmion, we map in Fig.~5(b) the $\Delta\rho^{\rm{PHE,\, Hys}}_{yx}$ defined as difference calculated by subtracting $\rPHE$ with decreasing field from that with increasing field, which removes the $M$-induced PHE showing a significant contribution above $\Hc$ between 20--50 K. As described above, the in-plane skyrmion formation largely depends on the magnetic field history; namely, skyrmions tend to coexist with the ferromagnetic (helical) state in the increasing (decreasing) field process. That hysteretic behavior is presented as positive [blue part in Fig.~5(b)] or negative [red part in Fig.~5(b)] $\Delta\rho^{\rm{PHE,\, Hys}}_{yx}$, while there is no hysteretic signal in the other $T$-$H$ region. We note that the magnetic phase diagram determined by PHE is different from that of previous study \cite{Wilson} based on the magnetization measurement.

\begin{figure}
\begin{center}
\includegraphics*[width=8.5cm]{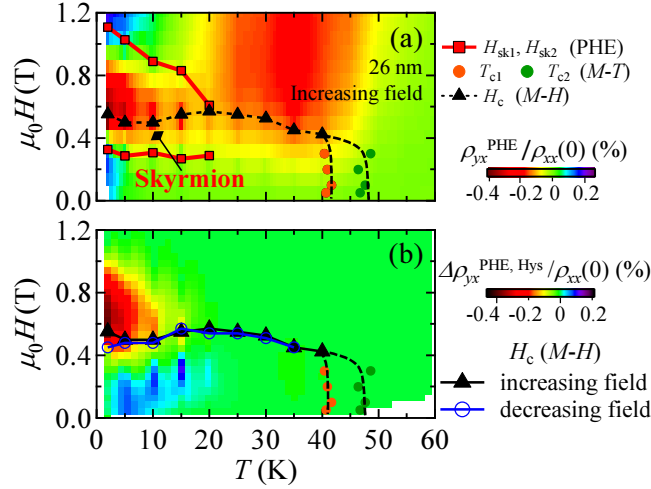}
\caption{(color online). Color maps of (a) planar Hall resistivity ($\rPHE$) normalized by longitudinal resistivity at zero field with increasing field and (b) $\Delta\rho^{\rm{PHE, Hys}}_{yx}/\rho_{xx}(0)$ defined by the hysteretic component of $\rPHE$ in the magnetic field scans [see Figs. 4(i)-(k)]. Squares represent $\Hsko$ and $\Hskt$. Solid triangles and open circles represent $\Hc$ with increasing field and decreasing field, respectively. Orange and green circles represent $\Tco$ and $\Tct$, which show the onsets of the long-range order and the intermediate regime (chiral spin short-range order), respectively. }
\end{center}
\end{figure}

Finally, we discuss the thickness ($t$) dependence of planar Hall signal [Fig. 6]. At low temperatures, where we demonstrate the in-plane skyrmion formation, a polarized neutron reflectometry study \cite{Wilson_3} has proposed a helicoidal state. The helicoidal state proposed in Ref. 27 shows discrete changes in its helix turns with a magnetic field variation. When the sample thickness is $n\lambda\le t<(n+1)\lambda$, where $\lambda$ is helical period, the helicoidal state with $n$-turns is realized. With application of the magnetic field,  the turns would be discretely unwound. If we assume the large kink in PHE [e.g. see Fig. 4(i)] originates from the helicoidal structure, namely the discrete change in the number of turns, additional kink would appear in a thicker film. Figure 6 shows that the PHE signals in 26 and 50-nm thick films. Even if we increase the thickness twice, the overall feature remains unchanged ; this is inconsistent with the model of the helical structure formation, but supports the present interpretation, i.e. the in-plane skyrmion formation.

\begin{figure}
\begin{center}
\includegraphics[width=8.5cm]{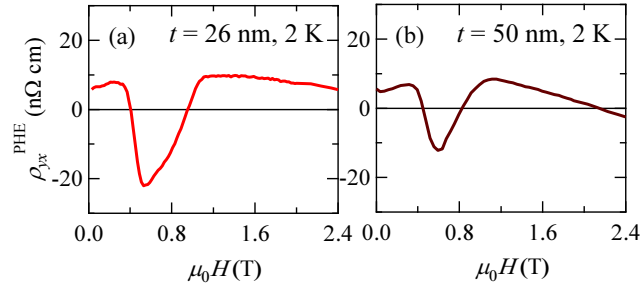}
\caption{(color online). Magnetic-field dependence of $\rho_{yx}^{\rm{PHE}}$ at 2K in (a) 26 nm film and (b) 50 nm film.}
\end{center}
\end{figure}

In conclusion, by measurements of PHE, we have revealed the formation of the in-plane skyrmions in the MnSi epitaxial thin film, which can hardly be detected by the conventional detection methods such as Lorentz TEM and topological Hall effect. PHE sensitively detects the 90$^{\circ}$-flop of the magnetic modulation associated with the skyrmion formation and destruction, showing the prominent stepwise anomaly in the skyrmion phase. 
We could determine the development of the respective magnetic texture in the MnSi film under the in-plane magnetic field, including the hysteretic formation of the in-plane skyrmions against the magnetic field change.
The uniaxial magnetic anisotropy due to the strain is likely the cause of the in-plane skyrmion formation at low temperatures.

\begin{acknowledgments}


The authors thank T. Ideue for enlightening discussions. This work is supported by JSPS through the Funding Program for World-Leading Innovative R\&D on Science and Technology (FIRST program), and Grant-in-Aids for Scientific Research (S) (No. 24224009 and No. 24226002) and for Young Scientists (Start-up) (No. 26886005).

\end{acknowledgments}

\end{document}